\documentstyle[prl,aps,twocolumn,epsf]{revtex}

\newcommand{\beq}{\begin{equation}}
\newcommand{\eeq}{\end{equation}}
\newcommand{\bea}{\begin{eqnarray}}
\newcommand{\eea}{\end{eqnarray}}
\newcommand{\eps}{\epsilon}
\newcommand{\veps}{\varepsilon}

\newcommand{\nn}{\nonumber}
\newcommand{\benn}{\begin{displaymath}}
\newcommand{\eenn}{\end{displaymath}}

\begin{document}

\title{\bf \LARGE 
Renormalization of the  Hartree--Fock--Bogoliubov 
Equations in the Case of a Zero Range Pairing Interaction }
\vspace{0.75cm}
\author{Aurel Bulgac and Yongle Yu}
\vspace{0.50cm}
\address{Department of Physics, University of
Washington, Seattle, WA 98195--1560, USA}
\maketitle

\today

\begin{abstract}

We introduce a natural and simple to implement regularization scheme of
the Hartree--Fock--Bogoliubov (HFB) equations with zero range 
pairing interaction. This renormalization scheme proves to be equivalent
to a simple energy cut--off with a position dependent running coupling constant.

\end{abstract}

\draft





More than forty years after the pioneering work of Bohr, Mottelson and Pines
\cite{pines} there is no need to reiterate again the relevance of pairing 
correlations in nuclei. It is well established  that nuclei are $s$--wave 
``superconductors'' in the so called weak coupling limit, when the pairing gap is
much smaller then the fermi energy $\Delta \ll \eps _F=\hbar^2k_F^2/2m$ 
($k_F$ is the fermi wave vector).  In this limit one can show that the 
rms radius of  the Cooper pair (in infinite matter)  significantly exceeds
the interparticle separation,  $\hbar^2k_F/m\Delta \gg 1/k_F$ \cite{mohit}, and
the radius of the nucleon--nucleon interaction as well. 
As in the case of the deuteron,  the details
of  the two--particle interaction at distances smaller or comparable with 
the interaction radius should be irrelevant and the bulk Cooper pair properties should 
be described basically by a single constant, derivable from a suitably chosen 
zero range interaction model.  (We shall not address here corrections beyond
the leading order, such as effective range effects.)
One encounters typically no insurmountable
difficulties in introducing a local HF (or Kohn--Sham)  Hamiltonian
$h(\bbox{r})$ \cite{hk,negele}.  If one can adopt the 
approximation of a zero range two--body interaction in the pairing channel as
well then  the HFB equations become
\bea
& &  [h (\bbox{r})  - \mu] u_i (\bbox{r})
     + \Delta (\bbox{r})  v_i (\bbox{r})
    = E_i u_i (\bbox{r}) , \label{eq:hfb0u}\\
& &  \Delta^* (\bbox{r}) u_i (\bbox{r})  -
    [ h (\bbox{r}) - \mu ] v_i (\bbox{r})
     = E_i v_i (\bbox{r}).
\label{eq:hfb0v}
\eea
Here $u_i (\bbox{r})$ and $v_i (\bbox{r})$ are the quasi--particle wave
functions,  $\Delta (\bbox{r})= \frac{\delta E_{gs}}{\delta \nu^*(\bbox{r})}$ is the local
pairing field, $\mu$ is the chemical potential,  $E_{gs}$ is the ground 
state energy  of the system and $ \nu(\bbox{r})$ is the  anomalous density.
In all the formulas presented here we shall not display the spin
degrees of freedom.  If one takes at face value
Eqs.  (\ref{eq:hfb0u},\ref{eq:hfb0v}) one can show that the diagonal
part of the anomalous density matrix $\nu (\bbox{r},\bbox{r})$ diverges,
since when $|\bbox{r}_1-\bbox{r}_2|\rightarrow 0$ the anomalous density
$\nu (\bbox{r}_1,\bbox{r}_2)$ has the singular behavior
\beq
\nu (\bbox{r}_1,\bbox{r}_2)=\sum _i v_i^*(\bbox{r}_1)u_i(\bbox{r}_2) \propto
\frac{ 1}{|\bbox{r}_1-\bbox{r}_2|},
\eeq
and the local pairing field
$\Delta (\bbox{R})$  cannot be defined \cite{ab,note1,bruun}.

In metals this type of singularity does not play a noticeable role,
because the summation over the single--particle states is cut--off at
energies of the order of the Debye energy $\omega_D \ll \veps_F$.
The single--particle density of
states is essentially constant in an energy window of width
${\cal{O}}(\omega_D)\ll \eps _F$ and the expression for the anomalous
density has only an infrared logarithmic divergence.  This logarithmic
divergence is due to states near the Fermi surface and has nothing to
do with the ultraviolet divergence due to states faraway from the
Fermi surface, which leads to the $1/|\bbox{r}_1-\bbox{r}_2|$
singularity discussed here. The infrared divergence leads to the
notorious non--analytical dependence of the gap on the coupling
constant, namely $\Delta = \omega _D\exp (-1/VN)$, where $V$ is the
strength of the interaction and $N$ is the single--particle density of
states at the Fermi energy $\veps_F$.

In nuclei and especially in very dilute fermionic atomic systems,
where $k_Fr_0\ll 1$ and $r_0$ is the radius of the interaction, there
is effectively no well defined cut--off and one needs to regularize
the theory. A finite range interaction will provide a natural cut--off
at single--particle energies of the order of $\veps _c\sim \hbar
^2/mr_0^2$, when the fast spatial oscillations of the quasi--particle
wave functions $u_i(\bbox {r}), v_i(\bbox{r})$ will render the nonlocal pairing
field $\Delta (\bbox{r}_1,\bbox{r}_2)$ ineffective. Even though the
presence of a finite range of the interaction in the pairing channel
formally removes the ultraviolet divergence of the gap, it is very
difficult to come to terms with the fact that a cut--off at an energy
of the order of $\hbar^2/mr_0^2$ could be the responsible for the
definition of the gap both in the case of regular nuclei and very
dilute nuclear matter as well. The characteristic depth of the 
nucleon--nucleon interaction potential, which is of the order of 
$\hbar^2/mr_0^2$, being the largest energy in the system, can 
be effectively considered to be infinite in the case of dilute systems. A well 
defined theoretical scheme for the calculation of a local pairing field, 
should lead to a converged result when only single--particle states 
near the Fermi surface are taken into account.

Most of the calculational schemes suggested so far for infinite
systems reduce to replacing a zero range potential by a
low energy expansion of the vacuum two--body scattering amplitude
\cite{mohit,blatt,yang,gorkov,randeria,george,hsu,khodel,fayans}. The
traditional approach in the calculations of finite nuclei consists
however in introducing a simple energy cut--off, while the pairing
field is computed by the means of a pseudo--zero--range interaction.
In this approach the effective range of the interaction is obviously
determined by the value of the energy cut--off and the two--body
coupling constant in the pairing channel is chosen accordingly\cite{henning}. 
Such a pure phenomenological approach lacks a solid theoretical underpinning
and always leaves the reader with a feeling that ``the dirt has been
swept under the rug''.  Another solution favored by other
practitioners is to use a finite range two--body interaction from the
outset, such as Gogny interaction \cite{ring}.  Besides the fact that
the ensuing HFB equations are much more difficult to solve
numerically, such an approach also lacks the elegance and transparency
of a local treatment and this seemingly simple recipe is indeed as
phenomenological in spirit as the treatment based on a
pseudo--zero--range interaction, with an explicit energy cut--off.
Moreover, in spite of the feeble arguments often put forward in favor
of a finite range interaction in HFB calculations, the only real
argument is the fact that the pairing field would otherwise diverge,
and there is no mean--field observable which would be noticeable
different in the case of a finite range interaction. 

The only attempt to implement a consistent regularization scheme for
finite systems that we are aware of is that of Ref. \cite{bruun}.  In
agreement with the analysis of Ref. \cite{ab} the authors of
Ref. \cite{bruun} conclude that in the case of a zero range two--body
interaction the anomalous density has a $1/|\bbox{r}_1-\bbox{r}_2|$
singularity. The regularization schemes for infinite homogeneous
systems amounts to subtracting a term proportional to $1/k^2$ in the 
gap equation  in momentum representation \cite{mohit}, which in coordinate 
representation  corresponds naturally to a $1/|\bbox{r}_1-\bbox{r}_2|$ 
term as well.  Since the divergence in the anomalous density 
$\nu(\bbox{r}_1,\bbox{r}_2)$ is due to large momenta and  thus short distances,
it is not surprising that  the character of the divergence is not affected by the size of
the system. Bruun {\it et al.} 
advocate the use of the following calculational procedure for the anomalous 
density. First of all one represents the anomalous density as \cite{typo}
\bea
& & \nu(\bbox{r}_1,\bbox{r}_2) =
\sum _{E_i>0}  \left [ v_i^*(\bbox{r}_1)u_i(\bbox{r}_2) +
\frac{\Delta(\bbox{r})}{2}
\frac{\psi _i^*(\bbox{r}_1) \psi _i(\bbox{r}_2)}{\mu -\veps_i}\right ] \nn \\
& & -\frac{\Delta(\bbox{r})}{2}
 G_0(\bbox{r}_1,\bbox{r}_2,\mu),
\label{eq:bruun} \\
& & [h(\bbox{r})-\veps_i]\psi _i(\bbox{r})=0,\\
& &  [\mu-h(\bbox{r}_1)]G_0(\bbox{r}_1,\bbox{r}_2,\mu)
=\delta(\bbox{r}_1-\bbox{r}_2) ,
\eea
where $\bbox{r}=(\bbox{r}_1+\bbox{r}_2)/2$.  One can easily justify this
subtraction scheme in infinite homogeneous matter, since
$v_i^*(\bbox{r}_1)u_i(\bbox{r}_2) =
\Delta \psi _i^*(\bbox{r}_1) \psi _i(\bbox{r}_2)/2\sqrt{(\veps_i-\mu)^2+\Delta^2}$.
In the limit $\bbox{r}_1\rightarrow\bbox{r}_2$ the sum over single--particle states
in Eq. (\ref{eq:bruun}) is converging now and one has only to extract
the regulated part of the propagator $G_0(\bbox{r}_1,\bbox{r}_2,\mu)$, 
using the pseudo--potential approach \cite{blatt}
\bea
& & \nu_{reg}(\bbox{r}):=
\sum _{E_i>0} \left [ v_i^*(\bbox{r})u_i(\bbox{r}) +
\frac{\Delta(\bbox{r})\psi _i^*(\bbox{r})
    \psi _i(\bbox{r})}{2(\mu-\veps_i)}\right ] \nn \\
& & -\frac{\Delta(\bbox{r})}{2}G_0^{reg}(\bbox{r},\mu), \label{eq:nureg}\\
& &G_0^{reg}(\bbox{r},\mu)= \lim _{\bbox{r}_1\rightarrow\bbox{r}_2}
G_0(\bbox{r}_1,\bbox{r}_2,\mu) +
\frac{m}{2\pi \hbar^2|\bbox{r}_1-\bbox{r}_2|}
\label{eq:greg}
\eea
obtaining for the local pairing field
\bea
& & \Delta(\bbox{r})= \frac{4\pi |a|\hbar ^2}{m}
\sum _{E_i>0} \left [ v_i^*(\bbox{r})u_i(\bbox{r}) +
\frac{\Delta(\bbox{r})\psi _i^*(\bbox{r})
    \psi _i(\bbox{r})}{2(\mu-\veps_i)} \right ] \nn \\
& & - \frac{4\pi| a|\hbar ^2}{m}
\frac{\Delta(\bbox{r})}{2}G_0^{reg}(\bbox{r},\mu), \label{eq:delreg}
\eea
where $a$ is the two--particle scattering length ($a<0$).
The renormalization procedure and the extraction of the
regulated part from various diverging quantities is completely
analogous to the familiar procedures in Quantum Field Theory, with the  
only difference that in this case everything is performed in coordinate space.
One literally "throws away" the diverging terms and retains the nonvanishing 
finite contributions.

The approach suggested in Ref. \cite{bruun} has however two, related,  
problems and as is formulated is applicable for systems in a harmonic
trap only and does not apply to atomic nuclei or other
self--sustaining systems.  First of all,  after the divergence has been
eliminated, the regulated expressions for the anomalous density and for the
pairing gap, Eqs. (\ref{eq:nureg},\ref{eq:delreg}), are defined entirely in
terms of states in a certain neighborhood of the fermi level, since the
corresponding sums converge rather quickly.  
Only when one can establish a one--to--one correspondence between the 
HFB terms  $v_i^*(\bbox{r})u_i(\bbox{r})$ and the corresponding  HF expressions 
$\Delta(\bbox{r})\psi _i^*(\bbox{r}) \psi _i(\bbox{r})/2(\mu-\veps_i)$ it is
clear how to evaluate Eqs. (\ref{eq:nureg},\ref{eq:delreg}). 
There is no   one--to--one correspondence for self--sustaining systems  
\cite{ab,jacek},  where sufficiently deep bound hole states lie in the 
continuum and where often there is no one--to--one correspondence 
between the HF and HFB spectra around the fermi level. 
In the case of nuclei very close to the nucleon drip lines the HFB spectra are continuous 
essentially everywhere, while the HF spectra are not, and the one--to--one 
correspondence between HFB and HF is absent. The second and the most 
difficult aspect of the approach suggested in  Ref. \cite{bruun}  however is 
the fact that it requires the determination of the regular part of the single--particle
Green  function $G_0^{reg}(\bbox{R},\mu)$, for which there is so far no clear
computational scheme in the case of an arbitrary self--consistent
field.  These two problems are to a large extent related, as only the
whole expressions (\ref{eq:nureg},\ref{eq:delreg}) are uniquely
defined, but not each separate part.

Our suggestion amounts to a simple to implement approach. First of all we
introduce an explicit energy cut--off $E_c$ in evaluating the anomalous
density. In this way we can evaluate separately the HFB and HF sums in Eqs. 
(\ref{eq:bruun},\ref{eq:nureg},\ref{eq:greg},\ref{eq:delreg}) irrespective of
the existence of the one--to--one correspondence discussed above.
The final result is independent of  $E_c$, if this is
chosen appropriately. Secondly, we remark that there is no compelling reason to
use the exact HF single--particle wave functions, energies and propagator in Eqs. 
(\ref{eq:bruun},\ref{eq:nureg},\ref{eq:greg},\ref{eq:delreg})
and in order to construct the regulator one can use a Thomas--Fermi 
approximation for the relevant quantities. Since the divergence has an ultraviolet 
character, the Thomas--Fermi approximation is particularly well suited \cite{tf}. 
Thus we arrive at the following relations
\bea
& & G_0(\bbox{r}_1,\bbox{r}_2,\mu-U(\bbox{r})) = -
   \frac{m \exp( ik_F(\bbox{r})|\bbox{r}_1-\bbox{r}_2|)}{
     2\pi\hbar^2|\bbox{r}_1-\bbox{r}_2|}      \nn\\
& & =  -\frac{m}{2\pi\hbar^2|\bbox{r}_1-\bbox{r}_2|}
    -\frac{ik_F(\bbox{r})m }{2\pi\hbar^2}
    +{\cal{O}}(|\bbox{r}_1-\bbox{r}_2|), \label{eq:g0reg} \\
& & \nu_{reg}(\bbox{r}):=  \nu_c(\bbox{r})
  +\frac{ i \Delta(\bbox{r})k_F(\bbox{r})m}{4\pi\hbar ^2}  \nn \\
& &   +\frac{\Delta(\bbox{r})}{4\pi^2}  \int _0^{k_c(\bbox{r})}k^2dk
   \displaystyle{ \frac{1}{ \mu -
       \displaystyle{ \frac{\hbar^2k^2}{2m}   }-U(\bbox{r} )+i\gamma }  }
 \label{eq:nu0reg}  \\
& &= \nu_c(\bbox{r}) \label{eq:nu1reg}  \\
& &   - \frac{ \Delta (\bbox{r}) m k_c(\bbox{r}) }{ 2\pi^2\hbar ^2 }
   \left \{
         1- \frac{ k_F(\bbox{r}) }{  2 k_c(\bbox{r}) }
          \ln \frac{ k_c(\bbox{r})+k_F(\bbox{r}) }{ k_c(\bbox{r})-k_F(\bbox{r}) }  
  \right \}    , \nn  \\
& &    \nu_c(\bbox{r}) =   \sum _{E_i \le E_c} v_i^*(\bbox{r})u_i(\bbox{r}),   \\
& &    h(\bbox{r})        = -\frac{\hbar^2\bbox{\nabla}^2}{2m}+U(\bbox{r}),  \\
& &    E_c                   =   \frac{\hbar^2k_c^2(\bbox{r})}{2m}+U(\bbox{r}) -\mu , \\
& &    \mu                    =  \frac{\hbar^2k_F^2(\bbox{r})}{2m}+ U(\bbox{r}),
\eea
where the cut--off energy $E_c$ is chosen sufficiently far away from
the Fermi level to insure that the rhs of Eqs. (\ref{eq:nu0reg},\ref{eq:nu1reg}) has
converged. As usual one has to take the limit $\gamma \rightarrow 0+$
at the end of the calculations. The local wave vector $k_F(\bbox{r})$
is real only in the physically allowed region of the fermi level,
where the regularized part of the propagator is imaginary. This
imaginary part of the regularized propagator is, naturally, exactly
canceled by the corresponding imaginary part of the momentum truncated
propagator in Eq. (\ref{eq:nu0reg}).  If the
Fermi momentum becomes imaginary
(outside nuclei for example) one can easily
show that $ \nu_{reg}(\bbox{r})$
is still real. The pairing field has thus the simple expression
\bea
& & \Delta(\bbox{r})= -g_{\mathit{eff}}(\bbox{r})\nu_c(\bbox{r})
=-g\nu_{reg}(\bbox{r})
 \label{eq:gapnreg} \\
& & \frac{1}{ g_{\mathit{eff}}(\bbox{r})}=
\frac{1}{g} \nn \\
&&  -\frac{m k_c(\bbox{r})}{2\pi^2\hbar ^2}
\left [ 1
  -\frac{k_F(\bbox{r})}{2 k_c(\bbox{r})}
\ln \frac{k_c(\bbox{r})+k_F(\bbox{r})}{k_c(\bbox{r})-k_F(\bbox{r}) },
    \right ] .   \label{eq:geff}
\eea
where $g=4\pi\hbar^2a/m$. Surprisingly, these relations look very much like 
a simple position or density dependent renormalization of the coupling constant.  
For a typical nuclear potential which monotonically increases with the radial 
coordinate ($dU(\bbox{r})/dr >0$) one can easily show that 
$ d g_{\mathit{eff}}(\bbox{r})/dr >0$, thus the effective pairing interaction
is stronger inside than outside nuclear
matter (remember $g<0$). This is stark contrast with the behavior one would
get using the popular energy cut--off of a $g\delta(\bbox{r}_1-\bbox{r}_2)$ 
interaction, namely the vacuum renormalization scheme \cite{henning}. 
In this case the effective coupling constant is  
$ g_{vac}(\bbox{r})=g/[1 -gmk_c(\bbox{r})/2\pi^2\hbar^2]$ and one can then easily
show that $d  g_{vac}(\bbox{r})/dr <0$ if $dU(\bbox{r})/dr <0$.

It is instructive to apply this recipe to the case of infinite homogeneous
matter.  After a few simple manipulations one can show that the equation for
the gap reads
\bea
& & \frac{1}{k_F}\int _0^{k_c}dk
\frac{k
^2}{\sqrt{ (k^2-k_F^2)^2+k_P^4}} \nn \\
& & =
\frac{\pi}{2k_F|a|}
 \left [
  1 +\frac{2k_c|a|}{\pi} -\frac{k_F|a|}{\pi}\ln\frac{k_c+k_F}{k_c-k_F}
 \right ]
 \label{eq:gap2} ,
\eea
where $k_P^2=2m\Delta /\hbar^2$.  Using the methods described in Refs.
\cite{mohit,yang,gorkov,randeria,george,hsu} one would not get
the term with the log--function. The technical reason is that we used
$\Delta/(\veps_i-\mu)$ instead of $\Delta/\veps_i$ in Eqs.
(\ref{eq:bruun},\ref{eq:nureg},\ref{eq:delreg},\ref{eq:nu0reg})
respectively \cite{note2}, which enhances the convergence of the
corresponding sums or integrals discussed above.  Parametrically we
are allowed to make such a substitution as long as $|k_Fa|\ll 1$,
otherwise one should consider effective range corrections and higher
partial waves.  Even though the momentum cut--off $k_c$ appears
explicitly here, once this momentum cut--off is sufficiently large,
there is no dependence of the gap on the cut--off momentum. 

When evaluating the total energy of the system one has to be careful and 
calculate the expression \cite{george}
\beq
E_{gs}=  \int d^3r \left [ 
         \frac{\hbar^2}{2m} \tau_c(\bbox{r})
         -\Delta (\bbox{r})\nu_c(\bbox{r})
                                 \right ] + E_{pot},
\eeq
where $E_{pot}$ is the usual HF potential energy contribution, since the 
kinetic energy density $\tau_c(\bbox{r})=2\sum_{E\le E_c} |\bbox{\nabla} v_E(\bbox{r})|^2$
diverges in a similar fashion as  $\nu_c(\bbox{r})$ $E_c$, but $E_{gs}$ does not.
  
We have implemented this renormalization scheme for the pairing field
for both selfconsistent and non--selfconsistent calculations of
spherical nuclei. The normal and anomalous densities were computed
following the complex energy integration technique extensively used by
Fayans and his collaborators \cite{fayans}.  In order to illustrate the
convergence properties we present in Fig. 1 the neutron pairing field
$\Delta(\bbox{r})$ obtained as a
solution of the Eqs. (\ref{eq:hfb0u},\ref{eq:hfb0v},\ref{eq:gapnreg},\ref{eq:geff}) for a
range of cut--off energies $E_c$.  The calculations
were performed for a simple Woods--Saxon potential
with fixed parameters corresponding to a $^{110} Sn$
nucleus \cite{ben} and for a fixed value of the
chemical potential $\mu=-0.1$ MeV (essentially at the neutron drip line).  
The value of the bare coupling constant is
$g=-200 $ MeV$\cdot$ fm$^3$. The total energy converges equally
fast with $E_c$. The reasons why convergence is 
achieved for $E_c={\cal{O}}(\eps_F)$ and how one can improve on this aspect are
discussed in Ref. \cite{best}.

\begin{figure*}[h,t,b]

\begin{center}
\epsfxsize=7cm
\centerline{\epsffile{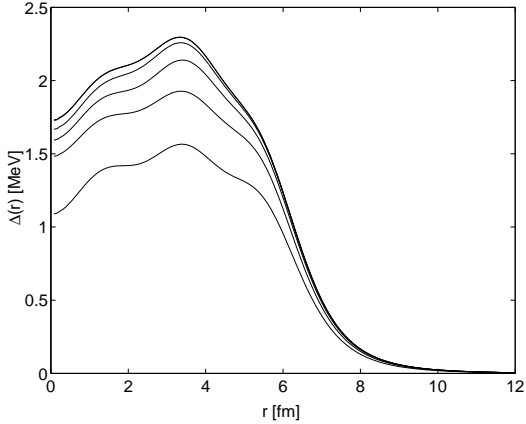}}
\end{center}

\caption{ The neutron pairing field (\ref{eq:gapnreg}) as a function of
the radial coordinate and of the cut--off energy $E_c$. Upward various curves
correspond to $E_c=20, \; 30, \; 35, \; 40, \; 45$ and 50 MeV respectively. On the scale of
the figure the last two curves are indistinguishable.}

\label{fig:fig1}

\end{figure*}

In conclusion, we have presented a renormalization procedure for the
HFB equations in the case of zero range pairing interaction, which is
easy to implement for any type of finite or infinite systems and which
converges very fast as well.  A very interesting feature of this approach
is its similarity with a density dependence of the pairing
interaction. The numerical implementation of the present renormalization scheme is
straightforward and amounts to very small changes of the existing
codes.

We thank DoE for financial support, G.F. Bertsch and J. Dobaczewski
for discussions and G. Bruun for drawing our attention to our initial
misinterpretation of the symbol $E_\eta^0$ in Eq. (27) of Ref. \cite{bruun}.
The very warm hospitality of N. Takigawa in Sendai
and the financial support of JSPS were very helpful  while AB was writing 
the final version of this work.



\begin{thebibliography}{99}
\bibitem{note1} The notations used in Ref. \cite{ab} are slightly
different from those commonly used in literature, which we try to
follow in this work.
\bibitem{tf} Since the Thomas--Fermi  approximation is used for estimating the
regulator and not the anomalous density or the pairing gap, no observables are
affected.
\bibitem{note2} The authors of Ref. \cite{bruun} state that previous
authors, in particular those of Ref. \cite{randeria}, have used
$1/(\veps_i -\mu)$ and not $1/\veps_i$ as a subtraction scheme. 
Such a replacement is accurate up to effective range
corrections only.
\end{thebibliography}
\end{document}